# Semantic Modeling and Retrieval of Dance Video Annotations

BALAKRISHNAN RAMADOSS[1]
KANNAN RAJKUMAR[2]

Department of Computer Applications
National Institute of Technology
Tiruchirappalli 620015, TN, India
[1]`brama@nitt.edu`
[2]`rajkumarkannan@yahoo.co.in`

**Abstract**. Dance video is one of the important types of narrative videos with semantic rich content. This paper proposes a new meta model, Dance Video Content Model (*DVCM*) to represent the expressive semantics of the dance videos at multiple granularity levels. The *DVCM* is designed based on the concepts such as video, shot, segment, event and object, which are the components of MPEG-7 MDS. This paper introduces a new relationship type called *Temporal Semantic Relationship* to infer the semantic relationships between the dance video objects. Inverted file based index is created to reduce the search time of the dance queries. The effectiveness of containment queries using precision and recall is depicted.

## 1. Introduction

Dances are archived in many ways. In ancient times, dancers passed the knowledge of the dance verbally to the following generations [13]. However, such knowledge was limited to the memory of the dancers and many dance productions have been lost due to this fact.

Traditionally, dance notations are used to archive the choreographies and to resurrect the dance production by the dancers and the choreographers. The two most common dance notations that succeed are Labanotation [2] and Banesh [9], which are developed by Rodolph Laban and Rudolf Banesh respectively. These two notational systems provide an abstract view of a three dimensional dancer in space. The fundamental problem with these systems is that very few people understand these notations and are experts in this field. The notation process involves the expert to visualize the representation of the notation and to translate the notation into the corresponding movement to the dancers.

Recording mediums such as CD, VCD, DVD and VHS, provide a quicker and easier way of acquiring and storing the dance performance. However, searching these collections of digital data is tedious [4] because of the huge volume of data.

Modeling the different semantics of the dance, archived in any of the above three mediums, is a large step. Apart from modeling, the semantic querying facilities to these archives are needed for the dancers and the dance learners. So a system with these semantic modeling and querying facilities will aid the present and future generation of dance teachers and students with better dance training and learning methods. With such system, dance students can understand and learn the dance pieces, emotions to be expressed while performing the dance pieces and also culture-specific features. Hence, this kind of dance video information system is highly desirable and more useful to the nation, as a cultural heritage.

This paper proposes a new dance video data model, temporal semantic relationships, inverted file based index structure and a query processor. The dance video data model abstracts the different dance video semantics, offers dance learners and choreographers semantic querying capabilities and captures the video structure as shot, scene and compound scene. The video structure characterizes the dance steps of *Pallavi*, Anu *Pallavi, Saranam* and *Chores* [13] of a song. Some of the temporal operators are *follows*, *repeats*, *observe*, *perform_same* and *perform_different* steps.

The rest of the paper is organized as follows. Section 2 presents some related works on dance video data models. The basics of dance video semantics are explained in Section 3. The dance video data model for the dance annotations is introduced in Section 4. Section 5 discusses the modeling of the temporal relationships. Section 6 depicts the class design using *DVCM*. Section 7 and section 8 present the algorithms for containment query processing and experimental results respectively. Finally, Section 9 concludes the paper.

## 2. Related Work

This section briefly reviews some of the existing video modeling and annotation proposals and discusses their applicability to dance videos.

The Extended DISIMA [10] model expresses the events and concepts based on the spatiotemporal relationships among the salient objects. However, in dance video systems, the model has to consider not only salient objects, but all objects such as instruments, costume, background etc.

A graph based video model [11] records the appearance and disappearance of salient objects and stores the trajectory details efficiently without any redundancy. This model is more suitable for sports and traffic surveillance videos, in which the system keeps track of ball movements and human movements. But it is not suitable for dance videos where we do not track the movements of the dancers or any other objects.

Forouzan et al [6] provide a system to extract some of the low level features such as color, silhouette of the dancers and sound. Also, the spatiotemporal features of the dancers in Macedonian dance are manually annotated for further query processing. The main focus of their system is to study the historical changes of a dance that undergoes over centuries.

COSMOS7 [3] models the objects along with a set of events in which the objects participate, as well as events along with a set of objects and temporal relationships between the objects. This model represents the events at a higher level only like *speak, play, listen* and not at the level of actions, gestures and movements. Also, emotions expressed by the body parts of the dancers, are not considered.

## 3. Dance Video Semantics

Dance videos contain several low-level and high-level features and provide ample scope for the efficient semantic retrieval and dance video mining. Low level features of the dance videos, such as color, texture and sound [12], do not unfortunately provide any meaningful cue to the dance video model. But on the other hand, dance videos provide very interesting high level features that can be captured in the video data model. These features include:
- accompanying song
- dance types such as *solo, duet, trio* or *group dance*
- dynamics such as *slow, graceful* and *staccato* movements
- accompanying music

These features reveal the structural characteristics of the dance videos.

### 3.1 Song Granularity and Song Types

Accompanying song of a dance is an important structural feature of the dance video and can be used as a cue for grouping the dance steps together. Hence, dance is viewed as a collection of steps that are portrayed to the audience in temporal order, according to the lyrics of the song.

Basically, a song [13] is composed of parts such as *Pallavi*(PA), Anu *Pallavi*(AP), *Saranam*(SA) and *Chores*(CH). *Pallavi* is the introduction of the song and is usually one or two lines in the lyrics, occurring in the beginning of the song. Anu *Pallavi* is the additional introduction to the song. *Saranam* is a group of lines, usually 2 to 8 lines in the lyrics representing a stanza. The *Chores* is also few lines with some interesting pieces. Generally, song components are group of rhyming lines forming a unit in the song. However, some of these components may be optional as described in song types shortly. Every song component groups a set of dance steps corresponding to the lines of lyrics of the song. Therefore, the song granularity corresponding to the dance steps is Song, Song Components and Dance Steps.

The different types of song can be enumerated as a set of regular expressions [8]. The song used in dance video retrieval system can be any of the six types.
- (PA AP SA$^+$)
- (PA SA$^+$)
- (SA$^+$)
- (PA AP SA (CH SA)$^+$)

- (PA SA (CH SA)$^+$)
- (SA (CH SA)$^+$)

The first regular expression type (PA AP SA$^+$) specifies that a song can have dance steps for one *Pallavi*, one Anu *Pallavi* and one or more *Saranam*s. Here, the operator + denotes one or more occurrences of SA. Similarly, the regular expression for the remaining song types can also be illustrated.

### 3.2 Dance Video Granularity

Each shot contains one dance step either Bharathanatyam or a casual step for each dancer. That is, a shot can contain several dancers performing dance steps. Group of lines of steps are known as either, *Pallavi*, Anu *Pallavi*, *Saranam* or *Chores*. The scene is an abstraction of any of the four song components. Similarly, compound scene is the abstraction of a collection of song components, according to the song type defined as a regular expression representing a song. Finally, dance video is a collection of songs of a dance presentation or a movie.

### 4. Dance Video Content Model

The aim of multimedia content modeling is to build a high level model of the multimedia domain based on the content description requirements. The model applies notions form both Entity-Relationship and UML methodologies in modeling the dance video content. The proposed model integrates the structural characteristics of the video such as video, compound scene, scene and shots, abstracting the corresponding granularity of the song. Moreover, it considers the different dance semantics such as dancers, costume, dance steps, background etc, apart from the spatio-temporal relationships among the dancers. This section presents the *DVCM* that efficiently describes the dance steps of the songs.

The dance video is composed of one or more shots. Each shot represents a dance step. This model does not consider frames to represent the dance steps. Each step has starting position of the dancers, intermediate movements of body parts and the final position. So a dance step cannot be abstracted in a frame. Further, the dance video can be modeled as segments. For instance, the set of dance steps constituting a song can be called as segments. Segments can be decomposed into sub segments. Therefore, a shot is a segment at the bottom level.

A segment depicts one or more events. Here dance steps are events that occur in the dance video. Objects are part of the events and dancers are objects who participate in the events. Graphical representation of *DVCM* is depicted in Fig.1. The entity classes and relationships of the model are formally defined as below:

### 4.1 Objects

Object can be distinctly identified by studying the entities in a domain. A class is used to describe the set of objects which have the same characteristics. Examples of structural objects are video, shot and segment and semantic objects such as object and event.

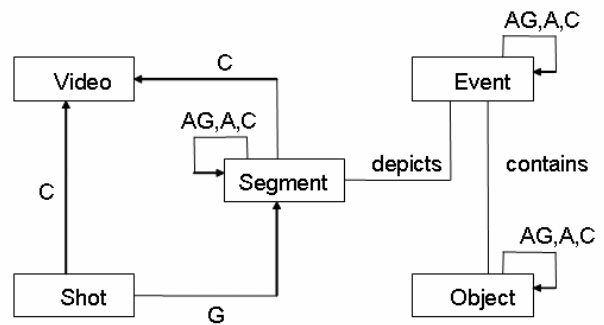

Fig.1.Graphical representation of *DVCM*.

### 4.2 Attributes

Attributes are descriptive properties of objects. There are three types of attributes.
- *Identifiers*: which uniquely identify objects like dancer ID.
- *Descriptors*: which describe objects such as age and sex of the dancers.
- *Temporal Attributes*: whose values change over time. There are two types of time elements [14]: *Transaction time* and *Existence time*. For instance, the attributes background sceneries and location are temporal attributes with existence time.

### 4.3 Relationships

An association among one or more objects is described as follows:
- *Generalization (G)*: It partitions entity classes into mutually exclusive sub classes. For example, in the dance video domain a shot is a segment.
- *Aggregation(AG)* and *Composition(C)*: Aggregation is called as *part-of* relationship. For instance, in dance video domain dancer is part of the dance step event.

- *Association (A)*: Association is a casual semantic relationship that relates objects, without exhibiting existence dependency. For example, in dance video domain, a dancer performs a dance step. Here, the association is *performs* which relates the semantic objects dancer and dance step.
- *Spatiotemporal relationships*: It is a relationship along the spatiotemporal dimension in which one object relates to another, spatially, temporally and spatiotemporally. For instance, dancerA is in front of dancerB facing each other and they move forward. Here, the spatio-temporal relationship are *in front of* and *move* (i.e. meet).

## 5. Temporal Semantic Relationships

The basic 13 temporal relationships for video modeling are defined by Allen [1]. Some more temporal relationships can be exclusively defined for dance videos, which are quite useful for query retrieval. These relationships correspond to the relationships among the objects that need not be stored, but can be inferred from the attribute values. Temporal relationships between the dancers, who are performing dance steps, capture temporal characteristics of the dancers. Some of the temporal characteristics of the dancers are:

- a dancer following a dance step of another dancer immediately.
- a dancer repeating a dance step of another some time later in the same song or another song.
- a dancer observing a dance step of another dancer.
- a dancer performing the same or different step of another dancer simultaneously.

Temporal relationships become the basis for any spatial relationship. For any two dancers who appear in a shot, the different semantic relationships abstracting the temporal characteristics can be defined as follows:

**Definition 5.1**. Let I be the time interval. Given two time intervals $I_1$ and $I_2$, $I_1$ is the *sub interval* of $I_2$, $I_1 \ll I_2$, if and only if $I_2.T_s \leq I_1.T_s$ and $I_2.T_e \geq I_1.T_e$, where $T_s$ and $T_e$ denote the start and end times of the time interval.

**Definition 5.2**. Let $D_i$ and $D_j$ be the dancers performing dance steps $S_1$ and $S_2$ in shots $SH_m$ and $SH_n$ of the scene $SC_k$ respectively. Let $I_1 = [T^i_s, T^i_e]$ and $I_2 = [T^j_s, T^j_e]$ be the time intervals associated with $D_i$ and $D_j$ respectively. Then $D_j$ *follows a step* of $D_i$, denoted as, $D_j \gg\rightarrow D_i$ iff
- $T^i_e = T^j_s$
- $SH_m.D_i.S_1 = SH_n.D_j.S_2$
- $I_1 \ll SC_k.I$ and
- $I_2 \ll SC_k.I$

**Definition 5.3**. Let $D_i$ and $D_j$ be the dancers performing dance steps $S_1$ and $S_2$ in the shots in shots $SH_m$ and $SH_n$ of the scene $SC_k$ respectively. Let $I_1 = [T^i_s, T^i_e]$ and $I_2 = [T^j_s, T^j_e]$ be the time intervals associated with $D_i$ and $D_j$ respectively. Then $D_j$ *repeats* a step of $D_i$, denoted as, $D_j \ll\rightarrow D_i$ iff
- $T^i_e < T^j_s$
- $SH_m.D_i.S_1 = SH_n.D_j.S_2$
- $I_1 \ll SC_k.I$ and
- $I_2 \ll SC_k.I$

Note: *Follows* and *Repeats* relationships are very interesting in a dance. Here, one dancer challenges the other with a dance step and the second dancer performs that dance step correctly.

**Definition 5.4.** Let $D_i$ and $D_j$ be the dancers performing dance steps in scene $SC_k$, associated with the intervals $I_1 = [T^i_s, T^i_e]$ and $I_2 = [T^j_s, T^j_e]$ respectively. Let SHS be the sequence of shots (in partial order) containing $D_i$ and $D_j$. Then $D_j$ *follows steps* of $D_i$, denoted as, $D_j \gg\gg\rightarrow D_i$ iff
- $T^i_e = T^j_s$
- Temporal relation $D_j \ll\rightarrow D_i$ holds for every shot s, where s ε SHS
- $I_1 \ll SC_k.I$ and
- $I_2 \ll SC_k.I$

**Definition 5.5.** $D_j$ *repeats steps* of $D_i$, denoted as, $D_j \ll\ll\rightarrow D_i$. It is similar to Definition 5.4, except the condition $T^i_e = T^j_s$ is replaced with $T^i_e < T^j_s$.

**Definition 5.6.** Let $D_i$ and $D_j$ be the dancers performing dance steps $S_1$ and $S_2$ in the shot $SH_k$ in the interval $I_1 = [T_s, T_e]$. Then $D_j$ *performs same step* together with $D_i$, denoted as, $D_j \equiv D_i$ iff
- $SH_k.D_i.S_1 = SH_k.D_j.S_2$
- $I_1 \ll SH_k.I$

**Definition 5.7.** Let $D_i$ and $D_j$ be the dancers performing dance steps $S_1$ and $S_2$ in the shot $SH_k$ in interval $I_1 = [T_s, T_e]$. Then $D_j$ *performs different step* together with $D_i$, denoted as $D_j \approx D_i$ iff
- $SH_k.D_i.S_1 \neq SH_k.D_j.S_2$
- $I_1 \ll SH_k.I$

**Definition 5.8.** Let $D_i$ and $D_j$ be the dancers performing dance steps $S_1$ and $S_2$ in the shot $SC_k$ in the interval $I_1 = [T_s, T_e]$. Let SHS be the sequence of shots (in partial order) containing $D_i$ and $D_j$. Then $D_j$ *performs same steps together* with $D_i$, denoted as, $D_j \equiv\equiv D_i$ iff
- $D_j \equiv D_i$ holds for each shot s, where s ε SHS
- $I_1 \ll SC_k.I$

**Definition 5.9.** Let $D_i$ and $D_j$ be the dancers performing dance steps $S_1$ and $S_2$ in the shot $SC_k$ in the interval $I_1 = [T_s, T_e]$. Let SHS be the sequence of shots(in partial order) containing $D_i$ and $D_j$. Then $D_j$ *performs different steps together* with $D_i$, denoted as, $D_j \approx\approx D_i$ iff
- $D_j \approx D_i$ holds for each shot s, where s ε SHS
- $I_1 \ll SC_k.I$

**Definition 5.10.** Let $D_i$ and $D_j$ be the dancers in a shot SH of the scene $SC_k$ in the interval $I_1 = [T_s, T_e]$. Let S be the dance step performed by $D_j$. Then $D_i$ *observes* step of $D_j$, denoted as, $D_j \Delta D_i$ iff
- $SH.D_i.S = NIL$ and $I_1 \ll SC_k.I$

## 6. Class Design based of DVCM

The following set of classes is designed in order to represent the different object types. They include Video, Compound Scene, Scene, Shot, Song, Musician, Background, Costume, Dancer, Dance Step, SpatialTriplet, Peytham(PY), Adavu(AD), Ashamyutha Hashtam(ASHA), Shamyutha Hashtam(SHA) and Casual Step(CS). PY is a Bharathanatyam step that is rendered with head, eye, eyebrow, nose, lips, neck, chest and sides only. AD is a Bharathanatyam step that is performed with legs and hands. ASHA is performed using fingers of either hand, while SHA with fingers of both hands simultaneously.

The class Video has attributes ID, life span, date of recording, video description and compound scene sequence IDs. The class Compound Scene includes attributes such as ID, videoID, songID, description and sequence of sceneIDs. The class Song has attributes such as ID, name of the song, lyrics and musicianID. The Musician class has ID, name, address, sex and phone attributes.

The class Scene includes the following attributes: ID, life span, compound sceneID, backgroundID, costumeID and sequence of shotIDs. Also, it has mapping $oc: O_i \rightarrow CO$ which maps each dancer to a set of costumes. The class Background has ID, background name, location and description attributes. The class Costume has ID, name and description attributes.

The class Shot has the following attributes: ID, life span, sceneID, set of dancerIDs, dance eventID, mapping $oa: O_i \rightarrow S_i$ which maps each dancer to a dance step and description. The class Dance event has the attributes step type, posture, reflexion, spatial triplet list and a mapping $oi: O_i \rightarrow IN_i$ which maps each dancer to an instrument. The class Instrument has ID, instrument name and description attributes. Posture is an enumeration of front, left, right, back side. Similarly, Reflexion is also an enumeration with values sad, happy, delighted and excited. Spatial Triple is the list with values dancer1, dancer2 and spatial relation between them.

The class Peytham has ID, peytham name and movement attributes. Similarly the class Ashamyutha Hashtam has ID, name and movement attributes. The classes Adavu and Shamyutha Hashtam have attributes: ID, name and movement of two body parts.

## 7. Dance Video Query Processing

Semantic queries that are related to dance videos can be classified into four types as below:

**Containment Queries:** In containment queries, a dance learner is interested to obtain video shots in which a dancer performs a step, a particular body part is used, a specific mood is reflected or an instrument is kept by a dancer and so on. For instance, a dance learner may submit a query, *Give me all video shots in which dancer A performs step S*, in order to learn the specific step S.

**Temporal Queries:** Temporal queries involve temporal relationships among the dancers. Temporal operators can be selected from Allen's algebra and our temporal semantic relationships, in order to facilitate the search and learning. For example a query, *Give me all video shots in which a step S done by dancer A is repeated by dancer B*. From this, one can conclude that this dance step may represent a *competition dance*, or is a repeated step which is an important one in a dance.

**Spatial Queries:** In these queries, choreographers and learners can express directional relationships between the dancers who perform same or different steps. These queries may be useful when the users want to know the spatial arrangement between two dancers or group of dancers. For instance, a query that the user may submit to the dance video retrieval system could be: *Find all video shots in which dancer A is to the left of dancer B performing steps*.

**Spatiotemporal Queries:** In these types of queries, the dance learners are concerned with analyzing the spatiotemporal relationships between the dancers. Consider a query, *Give me all video shots in which dancer A observes dancer B who performs step S and B is to the left of A*. The temporal operator *observes* indicates dancer *A,* is simply watching the step of dancer *B* without performing any step.

The choreographers, teachers and students can query the system to find shots, scenes or compound scenes which contain the specified dancers(D), body parts(AD), posture(PO), mood(RX), costume(CO), instrument(IN), background(BG) or dance step(ST) types such as PY, AD, ASHA, SHA or CS without any regard to the temporal or spatial relations among the dancers. Dance video queries can be composed using conjunctive and disjunctive operators. Containment query components are classified into three types as shown below:
- Type1: D, AG, PO, RX, IN, BG, CO
- Type2: PY, AD, ASHA, SHA, CS
- Type3: DPY, DAD, DASHA, DSHA, DCS, DPO, DRX

Due to the page limitations, this paper considers only containment query types.

**Algorithm1**: Type1 and Type2 cqueries
**Input:** query *qc*, video granularity *vg*, *Song song*
**Output:** IDs of shots, scenes or compound scenes
1. **for** each ID of *qc*
2.   **for** each shot in *song*
3.     **if** *qc* exists in object record of *shot* then
4.       $cshots_i$ = $cshots_i$ U *shot*.ID
5.     **endif**
6.   **endfor**
7. **if** *vg* = scene or *vg* = compound scene then
8.   **for** each shot ID in cshots
9.     get the scene ID from object record of shot ID
10.     $cscenes_i$ = $cscenes_i$ U scene ID
11.   **endfor**
12. **endif**
13. **if** *vg* = compound scene then
14.   **for** each shot ID in cscenes
15.     get compound scene ID from object record of scene ID
16.     $ccscenes_i$ = $ccscenes_i$ U compound scene ID
17.   **endfor**
18. **endif**
19. **endfor**
20. $resultset_i$ is cshots, cscenes or ccscenes depending on *vg*
21. return *resultset*

**Algorithm2**: Type3 cqueries
**Input:** dancer *D*, dance step *S*, *Song song*
**Output:** IDs of shots, scenes or compound scenes
1. find shot IDs of *D*, say *SHOTS1* using Algorithm1
2. find shot IDs of *S,* say *SHOTS2* using Algorithm1
3. *cshots = SHOTS1* ∩ *SHOTS2*
4. filter shot IDs from cshots which do not have *D* performing *S*
5. process *vg* of scene and compound scene as in Algorithm1
6. return *resultset*

Fig.2.(a) & (b). Sequential Search Algorithms.

## 7.1 Sequential Search

Sequential search retrieves the required video granularity by searching through the shot objects for the specified video query. For example, the query "find all video shots in which dancer *a* performs step *s*" returns all shot IDs where *s* is choreographed by *a* in a song. Algorithms for Sequential search of the containment query are depicted in Fig.2.

## 7.2 Inverted File Index Search

The inverted file index [7] speeds up the computation of the containment queries, by eliminating the sequential scan of the object records. The inverted file is the collection of inverted lists where each list defines an entry or key and a set of values for the entry. The following inverted files are created to facilitate processing of the containment queries.
- Dancer inverted file (DA_IF)
- Agent inverted file (AG_IF)
- Posture inverted file (PO_IF)
- Reflexion inverted file (RX_IF)
- Instrument inverted file (IN_IF)
- Background inverted file (BG_IF)
- Costume inverted file (CO_IF)
- Step inverted file (ST_IF)

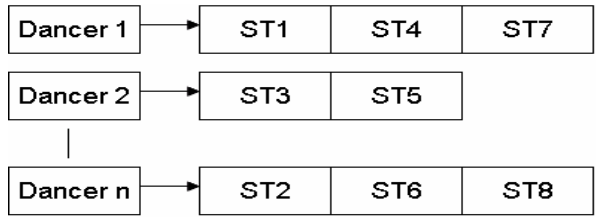

Fig.3. Dancer Inverted File.

**Algorithm3**. *Type-1* and *Type-2* containment queries
*Input:* query component *qc*, video granularity *vg*
*Output*: IDs of shots, scenes or compound scenes
1. **for** each ID of *qc*
2.   obtain a set of step IDs, say *STEPS*, from the respective inverted file if ID belongs to *Type-1* query or directly from the object record of Step in video data model if ID belongs to *Type-2* query
3.   **for** each step ID in *STEPS*
4.     obtain the corresponding shot IDs of the step ID, say *SHOTS*, from the step inverted file *ST_IF*
5.     $cshots_i = cshots_i \cup SHOTS$
6.   **endfor**
7.   process *vg* of scene and compound scene as in Algorithm1
8. **endfor**
9. return *resultset.*

**Algorithm4:** Type3 queries
**Input:** dancer *D*, dance step *S*, *Song song*
**Output:** IDs of shots, scenes or compound scenes
1. find shot IDs of *D*, say *SHOTS1* using Algorithm3
2. find shot IDs of *S,* say *SHOTS2* using Algorithm3
3. *cshots = SHOTS1 ∩ SHOTS2*
4. filter shot IDs from cshots which do not have *D* performing *S*
5. process *vg* of scene and compound scene as in Algorithm1
6. return *resultset*

Fig.4.(a) & (b). Inverted Index Search.

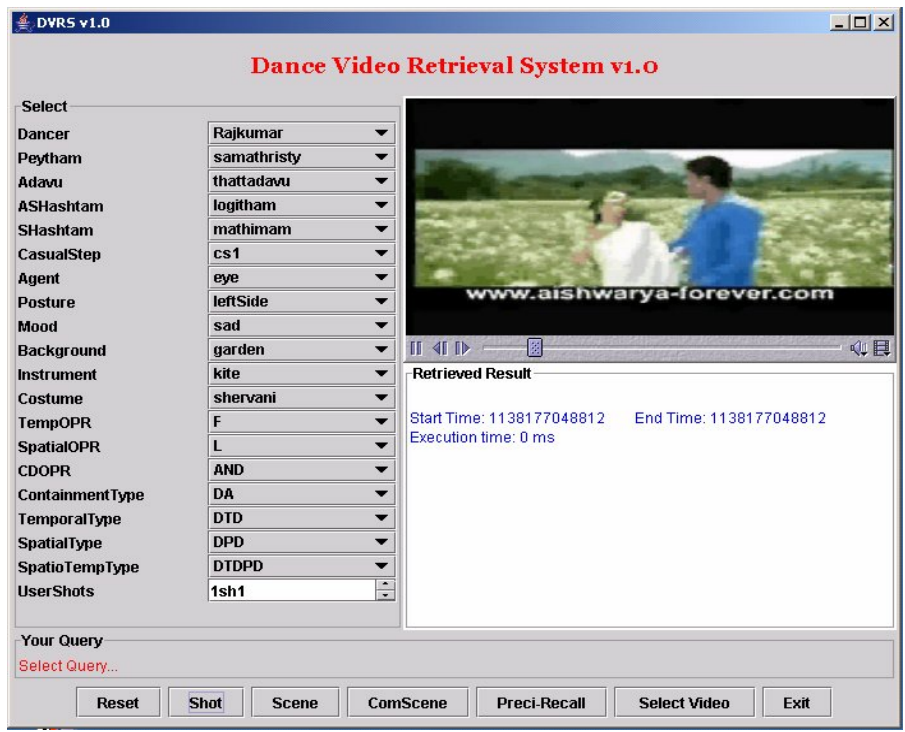

Fig.5. Screen shot of the Query Processor.

For instance, the structure of DA_IF is depicted in Fig.3. In the figure, each dancer is linked to an inverted list which contains a set of dance steps performed by the dancer in a song. Each node in the inverted list contains ID of the dance step. The inverted files AG_IF, PO_IF, RX_IF and IN_IF share the same structure as DA_IF. The BG_IF stores scene IDs for each background. That is, a set of steps in which the background appears is recorded in BG_IF. Similarly, CO_IF describes a set of scenes in which the specific costume is used by the dancers. The ST_IF describes a set of shot IDs for each dance step that belong to a song. Algorithms 3 and 4 (shown in Fig.4) present the query processing steps with inverted file index.

## 8. Performance

The experiments were run on Dell workstation with 512 MB DRAM, 80 GB HD under Windows XP environment. Fig.5 depicts the screen shot of the query processor which has been implemented using J2SE1.5 and JMF. The dance video users can select the query components such as dance steps and the query processor retrieves either shot, scene or compound scene based on the video granularity. Fig.6 shows the running time (RTIME) with shot groups 10, 100, 1000 and 10000 shots. As the number of shots increase, the Inverted file index greatly reduces the RTIME of the queries, compared to the sequential search. Due to lack of sufficiently annotated dance data, synthetic data have been generated representing the contents of the shots.

Further, the effectiveness of the containment queries (see Appendix-A) are studied using Precision and Recall [5] and are depicted in Table1. In this table, the average precision and recall works out to be 100% and 83.3% respectively. For the query Q5, the precision is less. The reason for this is that the system also retrieves the shots that match "joy", besides the romantic shots and hence the precision is less. Similarly, for the query Q2, the system retrieves the shots matching both left eye and right eye, thus making the precision less. For the remaining queries, the system achieves 100% because the retrieved shots are exactly same as the required shots.

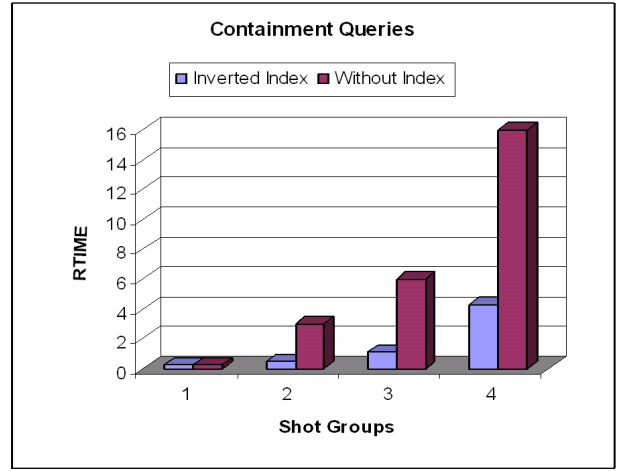

Fig.6. RTIME of Containment Queries.

| Query | Recall | Precision |
|-------|--------|-----------|
| Q1    | 100    | 100       |
| Q2    | 100    | 50        |
| Q3    | 100    | 100       |
| Q4    | 100    | 100       |
| Q5    | 100    | 66.66     |

Table1. Effectiveness of Containment Queries.

## 9. Conclusion

Dance videos possess rich semantics compared to other video types such as sports, news and movie. Dance video is multimedia comprising of textual, audio and video materials. This paper has proposed a dance video content model (DVCM) to incorporate the different dance video semantics such as dancer, steps, mood etc and video granularity. Set of temporal semantic relationships are introduced for spatial and temporal queries. The inverted file index speeds up the computation of the dance queries elegantly. The performance of containment queries under sequential and inverted file index search is also discussed.

## Acknowledgments

This work is fully funded by the University Grants Commission (UGC), Government of India Grant: XTFTNBD065.

**Appendix-A**

Q1:Show Anitha performing *Samathristy*

Q2:Show the dance steps of left eye performed by Lisa

Q3:Show Anitha performing dance steps denoting flowers

Q4:Show all dance steps of a song, *Kanna Varuvaaya*

Q5:Show all romantic dance steps